\documentclass{aa}

\usepackage{graphicx}
\usepackage{txfonts}

\usepackage{longtable}
\usepackage{lscape}

\usepackage{natbib}
\bibpunct{(}{)}{;}{a}{}{,}

\newcommand{\HS}{\mbox{\rm HS1603+3820}}
\newcommand{\zem}{\mbox{$z_{\rm em}$}}
\newcommand{\zabs}{\mbox{$z_{\rm abs}$}}
\newcommand{\Cf}{\mbox{$C_{\rm f}$}}
\newcommand{\kms}{\mbox{\rm km\thinspace s$^{-1}$}}
\newcommand{\cmsq}{\mbox{\rm cm$^{-2}$}}

\newcommand{\sh}{\mbox{$\sharp$}}
\newcommand{\aox}{\mbox{$\alpha_{ox}$}}

\newcommand{\Lya}{\mbox{{\rm Ly}\thinspace{$\alpha$}}}

\newcommand{\HI}{\ion{H}{i}}
\newcommand{\CII}{\ion{C}{ii}}
\newcommand{\CIV}{\ion{C}{iv}}
\newcommand{\SiII}{\ion{Si}{ii}}
\newcommand{\SiIII}{\ion{Si}{iii}}
\newcommand{\SiIV}{\ion{Si}{iv}}
\newcommand{\NV}{\ion{N}{v}}
\newcommand{\OI}{\ion{O}{i}}
\newcommand{\AlII}{\ion{Al}{ii}}
\newcommand{\AlIII}{\ion{Al}{iii}}
\newcommand{\FeII}{\ion{Fe}{ii}}
\newcommand{\FeIII}{\ion{Fe}{iii}}
\newcommand{\MgI}{\ion{Mg}{i}}
\newcommand{\MgII}{\ion{Mg}{ii}}

\newcommand{\err}{\mbox{$\pm$}}
\newcommand{\ttil}{\mbox{\char'176}}

\begin{document}

\title{Absorption spectrum of the quasar HS1603+3820\\
I. Observations and data analysis\thanks{Based on the observations
collected at the MMT, which is a joint facility of the Smithsonian
Institution and the University of Arizona. Some of the data presented
herein were obtained at the W.~M. Keck Observatory, which is operated
as a scientific partnership among the California Institute of
Technology, the University of California and the National Aeronautics
and Space Administration. The Observatory was made possible by the
generous financial support of the W.~M. Keck Foundation.}}

\author{A. Dobrzycki\inst{1}
        \and
        M. Nikolajuk\inst{2}
        \and
        J. Bechtold\inst{3}
        \and
        H. Ebeling\inst{4}
        \and
        B. Czerny\inst{5}
        \and
        A. R\'o\.za\'nska\inst{5}
}

\offprints{A. Dobrzycki}

\institute{ESO, Karl-Schwarzschild-Strasse 2, D-85748 Garching bei M\"unchen, Germany\\
\email{adam.dobrzycki@eso.org}
\and
Faculty of Physics, University of Bia{\l}ystok, Lipowa 41, PL-15424
Bia{\l}ystok, Poland\\
\email{mrk@alpha.uwb.edu.pl}
\and
Steward Observatory, University of Arizona, Tucson, AZ 85721, USA\\
\email{jbechtold@as.arizona.edu}
\and
Institute for Astronomy, University of Hawai`i, 2680 Woodlawn Drive,
Honolulu, HI 96822, USA\\
\email{ebeling@ifa.hawaii.edu}
\and
Copernicus Astronomical Center, Bartycka 18, PL-00716 Warszawa, Poland\\
\email{bcz,agata@camk.edu.pl}
}

\date{Received 22 February 2007 / accepted 27 September 2007}

\abstract{We present multi-wavelength observations of bright quasar \HS:
the optical data taken with the MMT and Keck telescopes, with the
40-50~\kms\ resolution, and X-ray data taken by the
\emph{Chandra X-ray Observatory} satellite.}
{The optical spectra contain a very large number of 
absorption lines from numerous heavy elements.
Our goal is to analyse those features to obtain constraints on the
properties of associated absorbers, to be used in modeling of the
quasar intrinsic flux and properties of the clouds.}
{We have determined the properties -- column densities and redshifts
  -- of the individual components. We derived the X-ray properties
of \HS\ and the optical-to-X-ray slope index \aox.}
{We found \aox\ of 1.70, which is on the high end of the
typical range for a radio quiet QSO.  We found 49
individual heavy element absorption clouds, which can be grouped into
eleven distinct systems. Absorbers from the associated
system which likely is the one spatially closest to the QSO
 show large \CIV\ to \HI\ column density ratio, reaching
$\sim$20.}
{Intrinsic X-ray properties of the quasar are typical. Determination
of column densities of ions (including hydrogen) gives a strong
foundation for modeling of the quasar ionising flux.}


\titlerunning{HS1603+3820 absorption spectrum. I}
\authorrunning{A. Dobrzycki et al.}

\maketitle

\section{Introduction}
\label{sec:intro}

\object{HS1603+3820}\ is a $\zem=2.54$ quasar discovered during the
Hamburg/CfA Bright Quasar Survey \citep{dobrzycki1996}. With $B=15.9$,
it is among the top few brightest quasars known in the $\zem=2-3$
redshift interval \citep{veron2006}. The most striking feature of \HS\
is, however, the extreme richness of its heavy element absorption
spectrum, and in particular the $\zabs\approx\zem$ absorption.

The absorption spectrum of \HS\ was first described in
\citet[][hereafter D99]{dobrzycki1999},
\defcitealias{dobrzycki1999}{D99} based on medium-resolution
(90-100~\kms) spectrum obtained in 1997 April with the MMT and the
Blue Channel spectrograph. Although in this spectrum many of the
absorption systems were blended, it was clear that the quasar held
great promise for the analysis of the properties of associated
absorbers.

The absorption systems which are spatially close to the quasar can be
used as probes for the intrinsic emission of the quasar itself, since
the conditions in the systems are undoubtedly heavily influenced by
the QSO flux \citep[e.g.][and references
therein]{crenshaw2003,gabel2005,scott2005}. Systems which contain
lines originating from various species and various ionisation levels
are particularly interesting, since they can provide independent way
of determining physical properties of the absorbing medium, which can
cast light at the intrinsic quasar continuum in far UV band as well as
the absorbing cloud formation.

With that as the principal goal, we initiated a project of
multi-wavelength observations of \HS, aiming at obtaining spectral
information at various wavelengths. In this paper we present the
results of our X-ray and optical observations of the quasar. In the
forthcoming paper (R\'o\.za\'nska et al.\ 2007, in preparation;
Paper~II) we use those observations to constrain models of the quasar
broad band continuum and the cloud confinement.

In the meantime, \citet[][hereafter referred to as M03 and M05,
respectively]{misawa2003,misawa2005} \defcitealias{misawa2003}{M03}
\defcitealias{misawa2005}{M05} observed \HS\ in 2002 March and 2003
July in very high resolution (6.7~\kms), using the Subaru telescope
with the High Dispersion Spectrograph. The primary goal of those
observations was to establish the nature of the absorbers (intervening
or associated), using the analysis of line time variability, partial
coverage and profile shapes. High resolution of their observations
enabled resolving many of the systems into a number of narrow
components. \citetalias{misawa2003,misawa2005} grouped the absorbers
into several systems, designated with the letters from A to H. In our
optical spectroscopy data we see all those systems and three
additional intervening systems.  For consistency, and to
enable easy comparisons, wherever possible we will use those
designations throughout this paper.

In a very recent paper, \citet[][hereafter M07]{misawa2007}
\defcitealias{misawa2007}{M07} expanded the analysis of absorber
variability, using new spectra from the Subaru and Hobby-Eberly
telescopes, and confirmed the finding from \citetalias{misawa2005}
that \CIV\ in system A is highly variable. This paper also described
one system (labeled I) which they did not have in their older
data. This system was seen in our data, and for consistency with
\citetalias{misawa2007} we rewrote parts of this paper to also use
this designation.

Our optical spectra are of lower resolution (FWHM of 40-50~\kms), but
they complement the Subaru data in two important areas: they have
higher signal-to-noise ratio (100-140 per pixel in the \CIV\ and \Lya\
regions) and they cover much larger wavelength range ($\sim$3700 to
$\sim10000$~\AA). The former reduces uncertainty in blended regions
containing thick lines and helps in finding weak lines, while the
latter allows analysing lines from species inaccessible in the Subaru
spectra.  The most important addition is detection and
analysis of hydrogen lines, not seen in the Misawa et al.'s spectra,
as well as detection of heavy-element lines other than \CIV\ in system
A. 

This paper is organised as follows. In Section~\ref{sec:obsxray} we
discuss the \emph{Chandra} X-ray observation of \HS. In
Section~\ref{sec:obsopt} we present our optical observations of \HS,
and we discuss the individual systems and compare our results with
those of Misawa et al., discussing the possible sources of
discrepancies. We comment on the absorber variability in
Section~\ref{sec:variability}. We summarise the results in
Section~\ref{sec:summary}.

\section{X-ray observations}
\label{sec:obsxray}

\HS\ was observed for 9.2~ks with the \emph{Chandra X-ray Observatory}
on 2002 November 29 (ObsID 4026), with the 128 row subarray of the
ACIS-S3 chip in VFAINT mode, with frame time of 0.4~sec. The
effective good time for the observation was 8.3~ksec.

The data were reduced using the tools available in the CIAO v.~3.3 and
SHERPA software packages\footnote{http://cxc.harvard.edu/ciao/}, using
Chandra calibration database (CALDB) v.~3.2.1. Because of large
calibration uncertainties for soft X-rays, the data with $E<0.5$~keV
were excluded from the analysis. We extracted source and background
spectra from the regions around the quasar using \emph{dmextract}. In
total, the net photon yield from the quasar was 145 photons, allowing
for most fundamental spectral fits only. We note that the ACIS pile-up
is not a problem, both because of the low source count rate and
because of the low frame time resulting from the use of the chip
subarray.

\begin{figure}
\centering
\resizebox{\hsize}{!}{\includegraphics{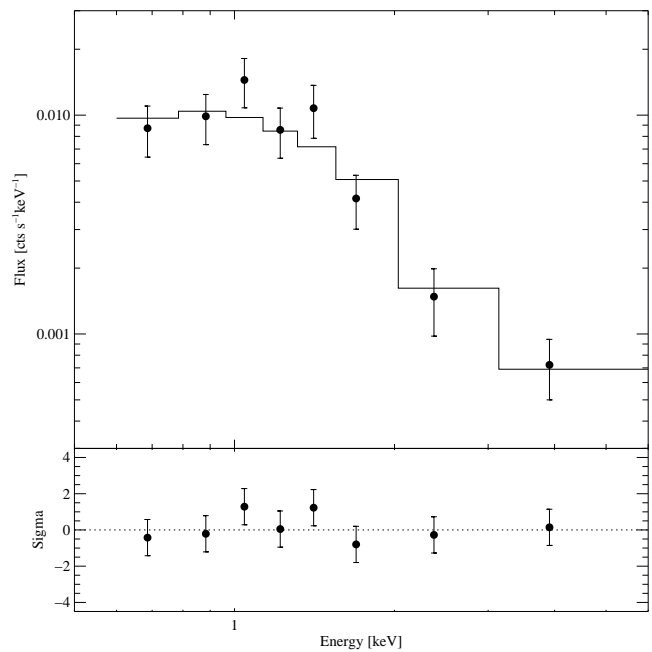}}
\caption{\emph{Chandra} ACIS-S3 X-ray spectrum of \HS\ and best fit
power law plus Galactic absorption. The bottom
panels shows the fit residuals.}
\label{fig:xspectrum}
\end{figure}

We performed spectral fits with SHERPA, applying the newest gain
corrections to the event list. We performed the fit assuming the
intrinsic quasar spectrum to be a power law, $A E^{-\gamma}$, with
fixed Galactic absorption towards \HS\ of
$N_H=1.3\times10^{21}$~\cmsq, as obtained with
COLDEN\footnote{http://cxc.harvard.edu/toolkit/colden.jsp}.
The best fit gave the photon spectral index of $\gamma=1.91\pm0.20$
and normalisation at (rest) 1~keV of $A=(2.43\pm0.79)\times 10^{-4}$
photons~cm$^{-2}$s$^{-1}$keV$^{-1}$. The slope of the spectrum is
within the typical range for quasars
\citep[e.g.][]{fiore1998,reeves2000}. The spectrum and the fit are
shown in Figure~\ref{fig:xspectrum}.

We also attempted fits in which we allowed for additional absorption
from the quasar, but there was no improvement in the fit.

Combination of the above result with the spectrophotometric data from
\citet{scott2000} yields the value of the optical-to-X-ray slope,
$\aox = -\log (f_{\nu,{\rm 2keV}} / f_{\nu,{\rm 2500A}})/\log(\nu_{\rm 2keV}/\nu_{\rm 2500A})$, 
of 1.70, which is at the high end of the typical parameter range at
$\zem\approx2.5$ \citep{bechtold2003}.

\section{Optical spectroscopy}
\label{sec:obsopt}

\HS\ has been observed by us with two instruments. On 2001 August 19
we used the ESI, Echelle Spectrograph and Imager \citep{sheinis2002},
on the Keck~II telescope, with the MIT-LL 2048$\times$4096 CCD, with
the 0.5\arcsec\ slit. Observing conditions were excellent throughout
the night, with clear sky and seeing of the order of 0.7\arcsec. The
combined exposure time was 4800~sec. The usable part of the spectrum
covered the $\sim$3900 to $\sim$10000~\AA\ wavelength range and had
the resolution of 45-50~\kms,  with the signal-to-noise ratio
peaking at $\sim$140 (per pixel) near the \Lya\ emission line,
gradually falling to $\sim$100 near the \CIV\ emission line and then
to $\sim$70 in the red end. 

On 2003 June 21 we used the Blue Channel Spectrograph on the MMT, in
the echelette/cross-dispersed mode, using the Lesser 2688$\times$512
``ccd35'' detector with the 1\arcsec\ slit. The observing conditions
were variable. Seeing became worse at the end of the night, reaching
$\sim$2~arcsec, and the conditions were adversely affected by the moon
age of $\sim0.75$ and smoke from forest fires. The combined exposure
time was 18,000~sec. The usable spectrum ranged from $\sim$3750 to
$\sim$7100~\AA\ and had the resolution of 40-45~\kms.  The
signal-to-noise ratio in this spectrum is in the 110--130 (per pixel)
range throughout the entire wavelength range. 

For both sets of observations, the initial reduction was done in the
standard way with IRAF. We note that despite several attempts, the
extracted MMT spectrum appears to show residual flux. Even lines that
are clearly saturated and which in the Keck spectrum have black cores,
do not reach zero intensity in their bottom in the MMT spectrum. We
elected to not correct manually for this effect, but we note that this
will likely result in underestimating the column density of the very
strong lines in the MMT spectrum.

After extracting the spectra from individual orders, we performed the
continuum fits to the spectra with the use of the FINDSL programme
\citep{aldcroft1994}.  The spectra were then normalised, rebinned to
common wavelength pixels and co-added, weighed with the variance.

Not surprisingly, the trickiest part of this part of the analysis were
the continuum fits.  The spectra are heavily absorbed bluewards of the
\Lya\ emission line. To fit continua in the uncertain areas, we
applied technique similar to the one described in
\citetalias{misawa2003}, i.e.\ utilising the spectrum shapes in the
unabsorbed orders. We note that this effect is less pronounced in our
case, since a single echelle order in our spectra -- which have lower
resolution than the Subaru spectra -- covers a larger wavelength
range, providing for more possibilities to ``anchor'' the continuum.

However, the continuum fits remain uncertain in two particularly
heavily blended areas. Comparison between continuum location in the
normalised spectra from Keck and MMT shows that there remains a
15-20\%\ uncertainty in those regions. In one case it is partly caused
by the fact that one of the ESI echelle orders ends inside the blended
region, forcing manual anchoring of the continuum fit. In the other
case it is the fact that the blended region is on the wing of the
emission line, which coincides with the peak sensitivity of the
echelle order; this results in very uncertain shape of the continuum
exactly where the technique of using shapes from unabsorbed regions
cannot be applied. We will return to this issue in
Section~\ref{subsubsec:sysa}.

We note that our continuum fits generally appear to agree well with
the Misawa et al.'s analysis of their data. One notable exception will
be discussed in more detail in Section~\ref{subsubsec:sysa}.

The next step in the analysis -- resolving and identifying absorption
lines -- was a very laborious process, since our spectra were heavily
blended. In order to avoid bias, we performed the fits blindly, i.e.\
we decided to not use the information from Misawa et al. As a
consequence, our results do differ quite significantly in some cases,
especially in the number of individual components into which some of
the systems were resolved. However, since the primary goal of our
analysis is to obtain information on the the column densities of as
many ions as possible in order to use them as constraints for modeling
of the quasar \emph{intrinsic} emission, the important information is
the total column density in the system, and not the number of the
individual components. In this regard, in the systems we have in
common -- primarily the vicinity of the \CIV\ emission line -- our
results do qualitatively agree with those of \citetalias{misawa2003}
and \citetalias{misawa2005}.

The line fits were done using VPGUESS/VPFIT
tools.\footnote{http://www.eso.org/{\ttil}jliske/vpguess/,
http://www.ast.cam.ac.uk/{\ttil}rfc/vpfit.html} This programme fits
multiple Voigt profiles, with  absorption redshift $\zabs$,
 column density $N$ and Doppler parameter $b$ as the
fitted parameters, and it is very well suited for analysis of blended
spectra. However, our spectra are exceptionally crowded and in some
cases we had to use simplifying assumptions in order to prevent the
programme from performing fits which were obviously non-physical. We
also had to apply some judgement calls. The cases where it was
necessary are mentioned in the following subsections. We note that the
absorption spectrum of \HS\ is so complex and so severely blended that
\citetalias{misawa2003} and \citetalias{misawa2005} had to resort to
similar approach even in the Subaru spectrum, which had markedly
higher resolution.

A large part of the work presented in \citetalias{misawa2005} was
devoted to the analysis of partial coverage of the QSO by the
absorbers. It is clear that at least some of the absorbers in the
vicinity of the quasar do not entirely cover the continuum emitting
region. In our analysis, we accounted for this effect only in the one
case where it was clearly necessary to obtain physically sound
parameter fits -- the thick \CIV\ lines in system A (see
Section~\ref{subsubsec:sysa}), which were clearly saturated but which
did not reach zero intensity in the line bottoms. We note that
\citetalias{misawa2005} presented arguments indicating that other
systems -- including those with $\zabs\ll\zem$ -- may also show
partial coverage. We will discuss this during the description of the
relevant systems.

The resolution of our spectra was slightly higher than the expected
values of the Doppler parameter for heavy element lines. High
resolution data from Subaru indicated that in many cases the metal
lines had $b$ as low as $\sim$5~\kms, while FWHM of 40-50~\kms\
translates (for Gaussian lines) to $b\approx 25-30$~\kms. On the other
hand, many lines have $b$ comparable to the spectrum resolution or
higher. In our analysis of the spectra we thus used the iterative
approach. We first freed all parameters in all fitted lines. Then for
all lines for which the resulting value of the Doppler parameter was
significantly below the resolution, we fixed $b$ at the value which
corresponded to the value of the parameter for other lines for which
it was well established  (using values for other metal lines
wherever it was possible and hydrogen lines in other cases),
 scaled with the square root of the element's atomic mass,
and re-run the fit. In the situation in which one or more lines had
inconsistent values of the Doppler parameter, we investigated the
possibility of having the line as a blend with other features.

It is relatively easy to fall into the trap of adding narrow lines in
order to improve the fits of large features. We avoided that issue by
adding lines only if it resulted in significant improvement of the
fit, and all cases when this was done were individually examined.

In total, we found 49 individual absorption clouds, which can be
grouped into eleven distinct ``systems.''  We stress that the systems
are defined based on their proximity in the spectrum, and the fact
that two absorbers are considered members of the same system does
\emph{not} in principle mean that they they are physically related.

In Tables~\ref{tab:sysa}-\ref{tab:sysek} we show line lists from all
systems detected in the spectra from Keck/ESI. In the tables, the
first column shows the line designation; the first component refers to
the separate cloud within the considered system, and the second part,
after the dot, is the ordinal number of individual line seen in this
cloud. The second column gives the ion identification (ID). The
following columns give the absorption redshift, the Doppler parameter,
column density and (in Table~\ref{tab:sysa} the covering factor,
\Cf, as defined in \citet{hamann1997}.  The last column gives (where
available) the line identification in \citetalias{misawa2005}. We mark
lines for which the derivation of parameters is made uncertain by
blending and/or by the fact that the line is saturated. The line lists
from the MMT/Blue Channel spectrum, as well as the entire Keck and MMT
spectra are available on the World Wide Web at
http://alpha.uwb.edu.pl/mrk/hs1603/.

In addition, in Table~\ref{tab:logNtot} we list the total column
densities of all ions detected in all systems and compare them to the
values found in the Subaru data.

The following section contains more detailed discussion of the
individual systems.

\subsection{Systems at $\zabs\approx\zem$}
\label{subsec:associated}

\begin{figure}
\centering
\resizebox{\hsize}{!}{\includegraphics{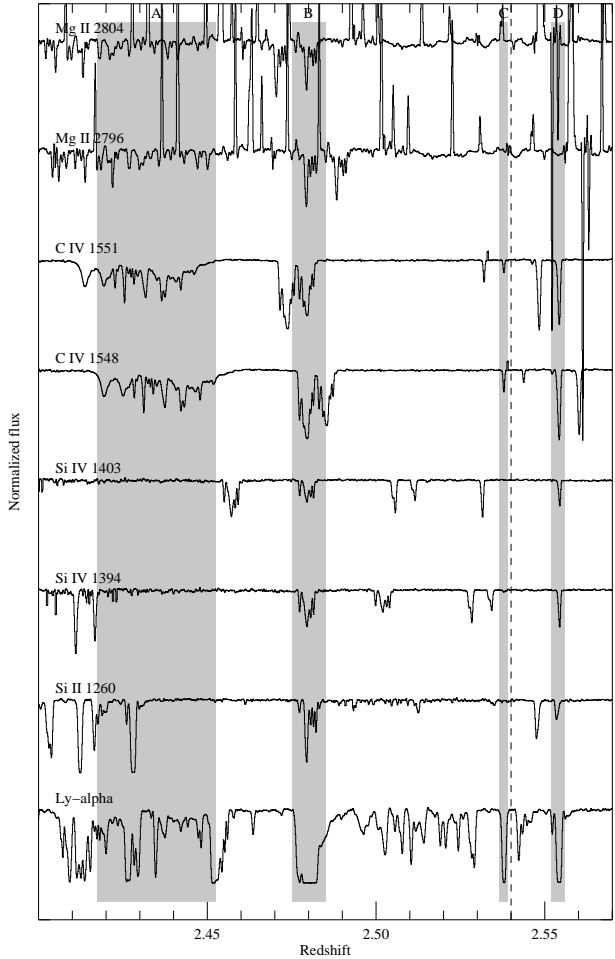}}
\caption{The fragments of the Keck spectrum of \HS\ plotted in the
same redshift range for several absorption features. Systems A-D are
shown as shaded areas. The \MgII\ lines are near the very red end of
the spectrum, and the ``emission'' features seen in their vicinity are
residuals of sky lines. The dotted vertical line shows the adopted
location of the QSO emission redshift.}
\label{fig:specad}
\end{figure}

Of the eleven distinct systems, four (A-D) are close to the emission
redshift of the quasar.  The ejection velocities range from
$\sim$10,000~\kms\ for system A to \emph{negative} (i.e.\
$\zabs>\zem$) $\sim$1,200~\kms\ for system D. 
Figure~\ref{fig:specad} shows the Keck spectrum of \HS\ in the
vicinity of the emission redshift of the quasar for several
absorbers. One can clearly see that some of the systems contain lines
from several species.

\subsubsection{System A, $\zabs=2.44$}
\label{subsubsec:sysa}

Five heavy-element absorbers were recognised in this system in the
medium-resolution data of \citetalias{dobrzycki1999}. At least one of
the components, at $\zabs=2.4189$ was identified as associated, since
the $\CIV(\lambda1548)/\CIV(\lambda1550)$ ratio was 1, indicating that
the cloud was optically thick, yet the lines did not reach zero
intensity in their bottoms, suggesting that the cloud in which the
lines originated did not obscure the entire continuum emitting source.

\citetalias{misawa2003} decomposed this system into seven individual
components, while \citetalias{misawa2005} increased this number to
fifteen systems. Their analysis also confirmed that the absorbers do
not appear to entirely cover the quasar, and they found indication
that this system is variable. These findings were confirmed and
expanded in the recent paper \citepalias{misawa2007}.

\begin{figure}
\centering
\resizebox{\hsize}{!}{\includegraphics{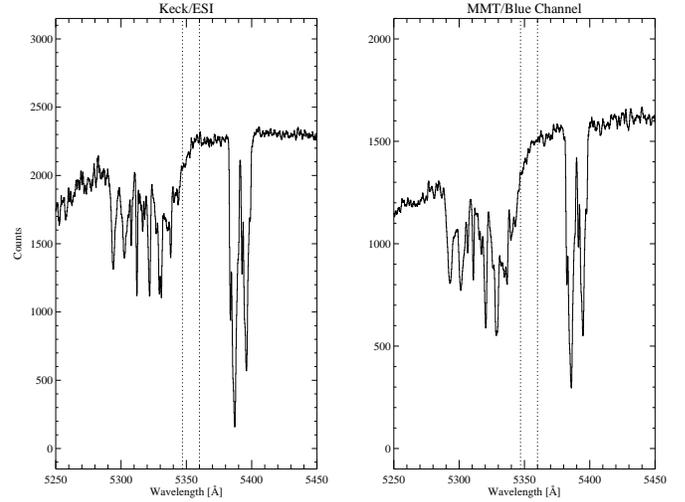}}
\caption{The sample raw spectra from Keck (left panel) and MMT (right
panel) as extracted from the echelle order, prior to continuum
fitting. The dotted lines show the $\sim$5347-5360 \AA\ region where
our continuum fit differs from Misawa et al. See text.}
\label{fig:dip}
\end{figure}

System A is unambiguously associated with \HS, and at the same time it
has a large relative velocity with respect to the quasar (ranging from
$\sim$7500 to $\sim$10,000~\kms), which means that it is almost
certainly located very close to the AGN central engine and its
properties are very likely determined by the radiation from the
QSO. This system is thus the ideal candidate for modeling of the
quasar continuum.

\begin{figure}
\centering
\resizebox{\hsize}{!}{\includegraphics{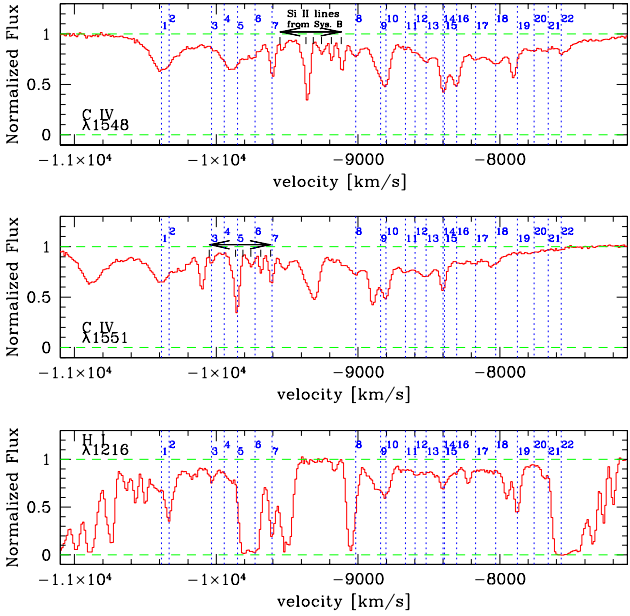}}
\caption{Absorption lines from System A. Top panel: \CIV($\lambda$1548),
middle panel: \CIV($\lambda$1551), bottom panel: \Lya. Vertical dotted
lines with numbers show the positions of individual lines as listed in
Table~\ref{tab:sysa}.}
\label{fig:sysa}
\end{figure}

In our analysis, we found twenty two components belonging to
system~A. The results are listed in Table~\ref{tab:sysa}; we show the
\Lya\ and \CIV\ lines on Figure~\ref{fig:sysa}. For the thick
absorbers in the blue end of the system we applied the correction for
partial covering using empirically determined covering factors (listed
in Table~\ref{tab:sysa}). We estimate the uncertainty on the covering
factor to be of the order of $\sim$10\%; the values agree
qualitatively with the values found in the Subaru data.

Our results for system A cannot be directly compared with the results
from Subaru, since at least four of the components seen in our spectra
(Nos.\ 19-22 in Table~\ref{tab:sysa}) result from the difference
between the continuum fits in both sets of data.

In their analysis, Misawa et al.\ (also in their very recent paper)
treated the wavelength region between $\sim$5345 and $\sim$5360~\AA\
as unabsorbed (see Fig.~4 in \citetalias{misawa2005} or Fig.~2 in
\citetalias{misawa2007}). However, this region clearly contains
absorption in both the Keck and MMT data. In Figure~\ref{fig:dip} we
show the \emph{raw} spectra of this region from both Keck and MMT
observations. In both cases one can clearly see that the continuum
does drop bluewards of $\sim$5360~\AA, and we had to fit lines in this
region in order to account for this drop. This was actually a
difficult task, since it was hard to resolve the dip into individual
systems. We primarily relied on the information from the \Lya\ region,
where strong, but heavily blended absorption was present. We found
that at least four systems (i.e.\ Nos.~19-22) with reasonable
properties are necessary to satisfactorily fit this region. However,
while we are confident that the absorbers are present in this region,
we have to treat their derived physical parameters as somewhat
uncertain.  We note that if the derived densities are taken
at face value, then the four systems contain of the order 20\%\ of all
\HI\ column density in system A, but only 2\%\ of the \CIV\ column
density; this would suggest that the presence of those clouds in the
vicinity of System A is coincidental and there is no physical relation
between them and the system absorbers. 

The other place in the system A area where there is a difference
between our analysis and the results from \citetalias{misawa2005} is
the $\sim$5308-5317~\AA\ region. This area contains the
\SiII($\lambda$1527) absorption from system~B (it is the feature near
$\lambda\approx5312$~\AA\ in Fig.~\ref{fig:sysa}). We obtained a very
good fit using eight lines, compared with 11 lines used in
\citetalias{misawa2005}. We note, however, that in the earlier
analysis \citetalias{misawa2003} also used 8 lines.

Even though it is a heavily blended region, our analysis was aided by
the fact that our spectra cover large wavelength ranges and contain
other \SiII\ lines from system B, which we could use to constrain the
properties of the $\lambda$1527 transition. Those lines clearly
indicated that this line is not as strong as described in
\citetalias{misawa2005}. However, this also indicated that this region
had to contain additional \CIV\ lines.  We added new \CIV\ doublets
which are labeled with 5, 6 and 7 in Fig.~\ref{fig:sysa} and
Table~\ref{tab:sysa}, and we also added line No.~3. Both spectra, and
especially the Keck spectrum, show that those doublets are
required. Additionally, we split \citetalias{misawa2005}'s line no. 1
into two lines -- our doublets Nos.~1 and 2. This significantly
improved our fit.

\begin{figure}
\centering
\resizebox{\hsize}{!}{\includegraphics{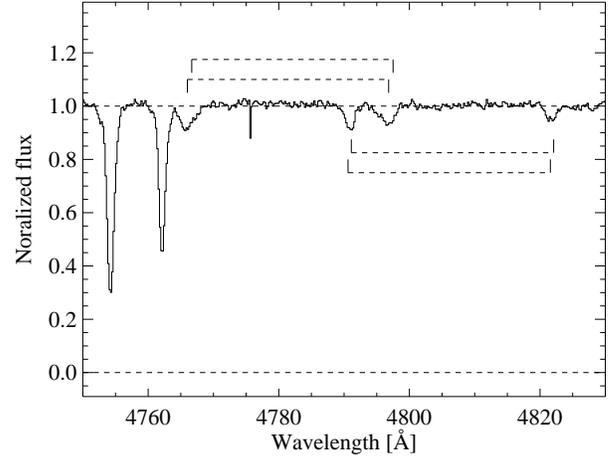}}
\caption{The \SiIV$\lambda\lambda$1394,1403 doublets from
system A.\label{fig:sysaSi}} 
\end{figure}

In System A, we also identified the \SiIV, \SiIII, and \NV\ lines. We
show the \SiIV\ lines in Fig.~\ref{fig:sysaSi}; the lines from the
latter two ions are in the heavily crowded \Lya\ forest region.  We
note that the \SiIV\ absorption was not seen in the Misawa et al.'s
data, even in the spectrum generated by coadding all the Subaru
spectra \citepalias{misawa2007}. This is clearly because of high
signal-to-noise ratio in our spectra.

We note that the components in system A have a rather large
value of the \CIV/\HI\ column density ratio, higher than 1 in
virtually all cases and reaching $\sim$20 in the extreme case. This is
seen in both Keck (Table~\ref{tab:sysa}) and MMT spectra, and has
strong implications for the modeling of the intrinsic spectrum of the
quasar (Paper~II). This effect is not seen in other systems, where
practically all components have $N_{\CIV}/N_{\HI}<1$.

 Our determination of the \CIV\ column density agrees
well with the value obtained in the Subaru data
(Table~\ref{tab:logNtot}). As mentioned earlier, none of the other
ions seen in our spectra was seen in the Misawa et al.'s data.

Finally, we would like to confirm the suspicion first expressed in
\citetalias{misawa2003} that the feature at 5527~\AA\ in the
\citetalias{dobrzycki1999} spectrum is \emph{not} a
\FeII($\lambda$1608) line at $\zabs=2.4367$. This feature is not
present in our Keck and MMT data, and a close examination of the raw
1997 data revealed that it was in fact an artifact caused by a cosmic
ray hit.

\subsubsection{System B, $\zabs=2.48$}
\label{subsubsec:sysb}

Even the medium resolution spectrum from \citetalias{dobrzycki1999}
clearly showed that this is a strong and complex system. The \Lya\
absorption was very broad, and its non-symmetric profile hinted that
it is in fact a blend of at least few lines.

\begin{figure}
\centering
\resizebox{\hsize}{!}{\includegraphics{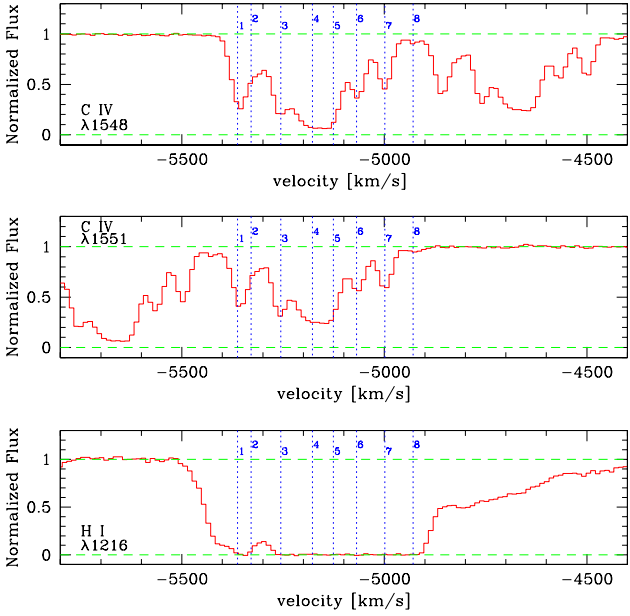}}
\caption{System B: the \CIV\ (upper panel) and \Lya\ (lower panel)
regions. Positions of lines from
Table~\ref{tab:sysb} are marked.}
\label{fig:sysb}
\end{figure}

The Subaru spectrum allowed this complex to be resolved into 18-19
separate components. As mentioned earlier, our spectra could be
satisfactorily fitted with only eight clouds. We show the vicinity of
\Lya\ and \CIV\ lines on Figure~\ref{fig:sysb}. The derived properties
of all lines are listed in Table~\ref{tab:sysb}. Note that this system
contains lines from as many as fourteen distinct ions.

It needs to be noted, however, that in few cases the line properties
are uncertain. In particular, the seemingly large column densities of
some of the \SiIII($\lambda$1206) lines (Nos.~2 and 6 in
Table~\ref{tab:sysb}) very likely originate from blending with the
hydrogen lines. Several other components are also likely affected by
features in the spectra; some of those features may be unidentified
lines, while some are very likely to be spectral artifacts. All those
cases are marked in Table~\ref{tab:sysb}.

Also, direct fit of \SiII\ (7.\SiII\ in Table~\ref{tab:sysb}) and
4.\FeII\ in the Keck spectrum resulted in unrealistically high column
densities. That was confirmed by visual inspection. The only
workaround was to perform the fit to those lines by trial-and-error,
then fix the column density for those components, and then refit the
whole spectrum again.

The 4.\OI\ lines near $\sim$4531\AA\ are affected by the \CIV\ doublet
from the J system (see below).

The reality of the 5.\AlIII, 5.\AlII\ and 5.\FeII\ lines is somewhat
uncertain. The programme fitted the lines and they were seen in visual
inspection of the spectrum, and we decided to keep them in
Table~\ref{tab:sysb}. However, we note that the lines themselves and
their vicinity are rather noisy.

Total ion column densities seen in our spectra agree well with those
seen in the Subaru data (Table~\ref{tab:logNtot}). Differences can be
seen for some ions, but they can be resolved when blends and suspected
blends are taken into account.

\subsubsection{System C, $\zabs=2.54$}
\label{subsubsec:sysc}

\begin{figure}
\centering
\resizebox{\hsize}{!}{\includegraphics{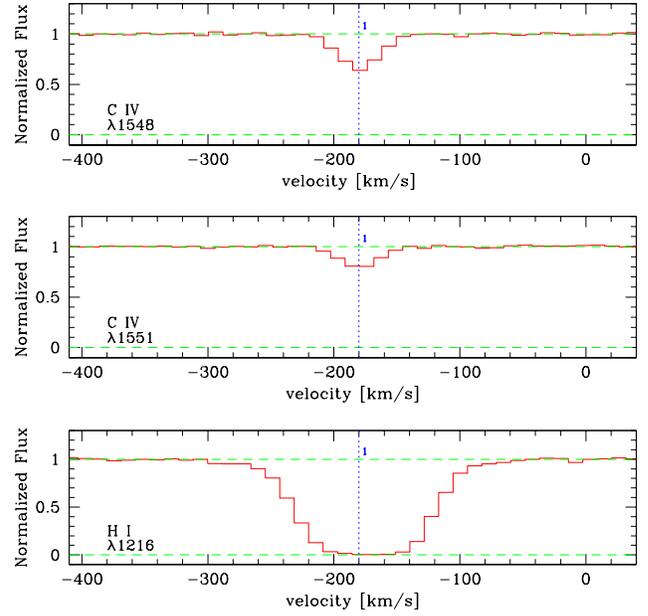}}
\caption{The \CIV\ (upper panel) and \Lya\ (lower panel) regions of
System C. Positions of lines detected in the
spectra are shown with dotted line and labeled with the ID from
Table~\ref{tab:syscd}.}
\label{fig:sysc}
\end{figure}

This system (see Figure~\ref{fig:sysc}) is very close to the emission
redshift of the quasar. In \citetalias{dobrzycki1999} it was
considered to be in-falling because of the QSO emission redshift of
2.51 adopted in this paper. However, the arguments presented in
\citetalias{misawa2003} favouring higher value of the emission
redshift, $\zem=2.54$, are convincing and we adopt this value in both
this paper and in Paper~II.

The derived properties of \CIV\ absorption (See Table~\ref{tab:syscd})
in this system agree well with the Subaru data. In our data we also
see the \Lya\ line. There is no significant absorption seen in other
lines, including \MgII.

\subsubsection{System D, $\zabs=2.55$}
\label{subsubsec:sysd}

\begin{figure}
\centering
\resizebox{\hsize}{!}{\includegraphics{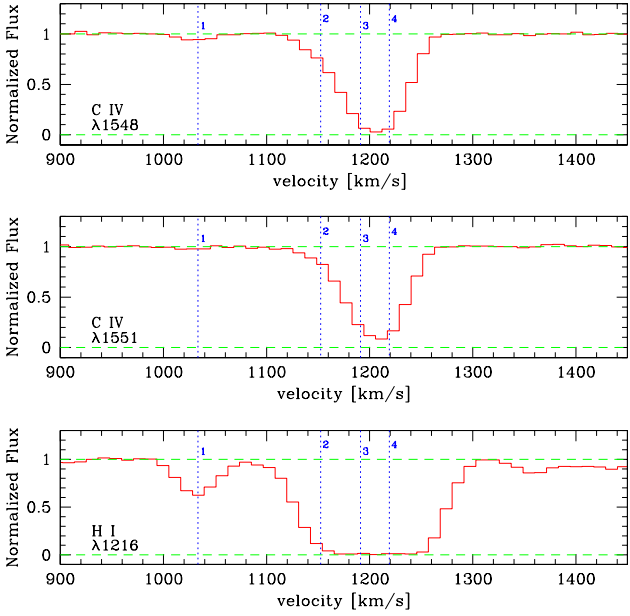}}
\caption{As in Figure~\ref{fig:sysc}, for System D.}
\label{fig:sysd}
\end{figure}

This system is undoubtedly a $\zabs>\zem$ system (see
Figure~\ref{fig:sysd}), even with the QSO emission redshift of
2.54. The natural explanation is that this system is infalling onto
the quasar.

Misawa et al.\ resolved this system into four separate components, and
we see the very same absorbers in our data. For all of them we
identify \Lya\ and \CIV\ lines.

The stronger two components (lines No.~3 and 4 in Fig.~\ref{fig:sysd}
and Tab.~\ref{tab:syscd}) also show \SiIV\ and \NV\ lines. We list
\NV\ only under component \#4, but it is very likely a blend of
components \#3 and \#4, since both lines in the doublet are broader
than what would be expected based on the value of the Doppler
parameter of the other lines. The nitrogen absorption is certainly
real, but the troughs are shallow and are visible only because of high
S/N of our spectra; resolving it into two components would be
meaningless.

In the case of the Keck spectrum, the fits of the \Lya\ lines \#3 and
\#4 were diverging when they were combined with lines from other ions.
We therefore fitted them separately and then fixed the values; this is
reflected in Table~\ref{tab:syscd}, where their values of $N$ and $b$
are listed as fixed.

We also note that the \HI\ line \#1 in the Keck spectrum shows
unusually low value of the Doppler parameter. However, this was
confirmed by the visual inspection of the spectrum: the line is
isolated but does appear to be very narrow. Since the corresponding
\CIV\ absorption is unambiguously identified, we have no doubts about
the reality of this line, and we attribute the unusually low measured
value of $b$ to noise fluke.

\subsection{Systems with $\zabs\ll\zem$}
\label{subsec:intervening}

The spectrum of \HS\ contains several other heavy-element systems at
$\zabs\ll\zem$. We identified seven of them in our spectra; the lists
of lines are shown in Table~\ref{tab:sysek}. Four of those systems: E
($\zabs=1.89$), F ($\zabs=1.97$), G ($\zabs=2.07$), and H
($\zabs=2.27$) were identified in the early Subaru spectra, while
system I ($\zabs=2.18$) was described in the recent paper by
\citetalias{misawa2007}.

In almost all cases, the properties derived from our spectra match
reasonably well with those from the Subaru data. The two exceptions
are the Nos.~3 and 4 \CIV\ lines from system F. In the first case
\citetalias{misawa2003} and \citetalias{misawa2005} have a doublet of
lines of similar strength separated by $\sim$7~\kms, which in our
spectrum is not deblended, and the second case is a weak line buried
in a nearby, stronger line.  The total column density derived
from our data does, however, agree well with the Subaru value
(Table~\ref{tab:logNtot}). 

\begin{figure}
\centering
\resizebox{\hsize}{!}{\includegraphics{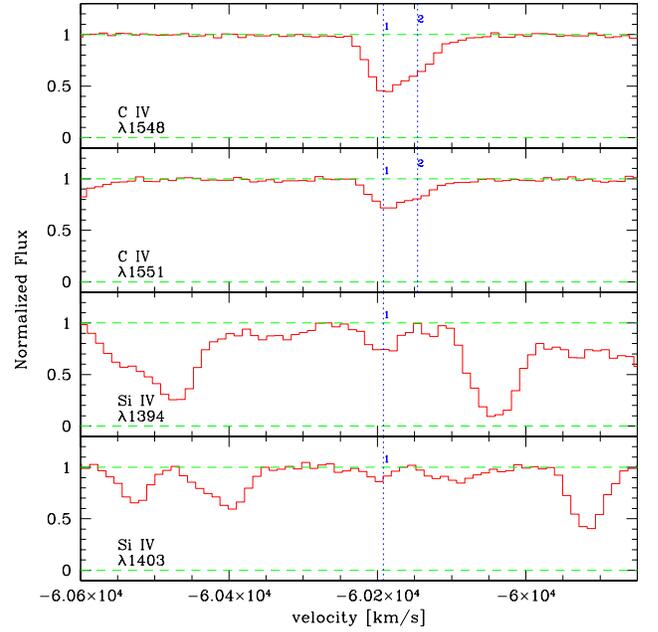}}
\caption{The \CIV\ (upper panel) and \SiIV\ (lower panel) regions for
System E. Markings as in Figure~\ref{fig:sysc}.} 
\label{fig:syse}
\end{figure}

Although the \SiIV($\lambda\lambda$1394,1403) lines in system E
(Fig.~\ref{fig:syse}) lie in the \Lya\ forest, they are clearly
identified in both Keck and MMT spectra. Existence of the second
\SiIV\ doublet is rather uncertain. We could not see it in the Keck
data, although it is perhaps present in the MMT spectrum.

The natural interpretation for the origin of the systems with
$\zabs\ll\zem$ is that they are intervening, i.e.\ they are caused by
objects which happen to be placed on the line of sight to the quasar
but are not related to it, although there are cases in the literature
where convincing arguments were presented that systems seemingly far
from the quasar could be associated
\citep[e.g.][]{hamann1997,richards1999}.  \citetalias{misawa2005}
indicated that systems E and F could be associated, since their
analysis seemed to show that some of the components in those systems
had partial coverage, although the lines did not show any
variability. In our analysis we obtained satisfactory profile fits
without the assumption of partial coverage. This accounts for the
difference between line parameters derived here and in the Subaru
spectra.  However, the total column densities agree well
(Table~\ref{tab:logNtot}). 

\begin{figure}
\centering
\resizebox{\hsize}{!}{\includegraphics{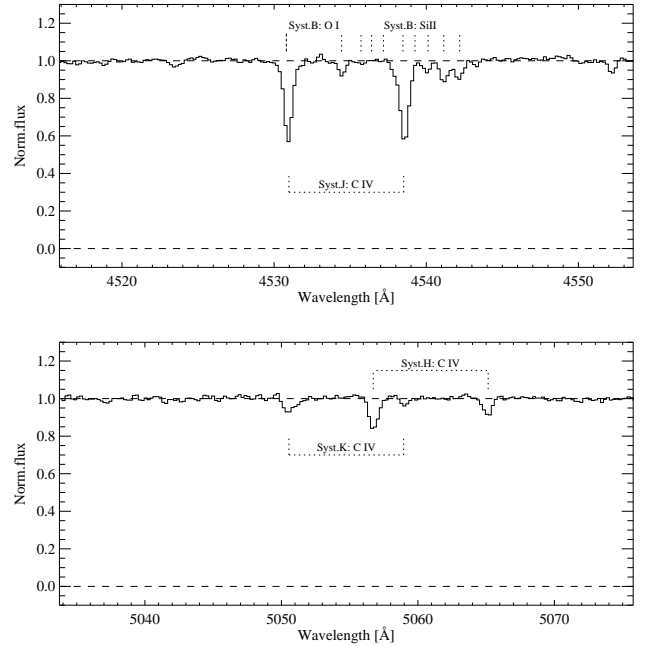}}
\caption{The \CIV\ regions for System J (upper panel) and K (lower
panel).}
\label{fig:sysjk}
\end{figure}

In our spectra, we also identify two systems which are not seen in the
Subaru spectra. We label them J ($z_{abs}=1.9266$) and K
($z_{abs}=2.2622$); the line properties are listed in
Table~\ref{tab:sysek}. We show the two systems in
Figure~\ref{fig:sysjk}.

System J requires a special comment. The \Lya\ line for this system
would be outside of the wavelength region covered by both MMT and Keck
spectra, and the only identified feature is the \CIV\ doublet. Before
we proceed, we note that it is somewhat surprising that the redder of
the two lines, at $\lambda_{\rm obs}\approx 4538$~\AA, is apparently
not identified in any of the Subaru spectra, since none of the line
lists seems to contain it. This line is unambiguously present in both
our spectra.

What makes this system somewhat uncertain is the fact that both those
lines coincide very well with the locations of what we expect to be
the strongest \OI($\lambda$1302) and \SiII($\lambda$1304) lines from
System B. However, there are indications that \CIV\ is in fact present
in this location. First, the spectra contain several other \SiII\
lines, and they constrain the column density of that ion quite well,
suggesting that \SiII($\lambda$1304) alone cannot account for all of
absorption in the redder of the two lines. Second, for \OI\ to be
responsible for all of the absorption in the bluer of the two lines,
at $\lambda_{\rm obs}\approx 4530$~\AA, it would require \OI\ column
density of $\sim10^{15}$\cmsq, which we consider implausible.

The above considerations lead us to believe that system J is real,
although the fact that both lines are blends makes its properties
rather uncertain.

Another of the previously unidentified systems is a system at
$\zabs=2.2622$, which we label K. It is clearly seen in both our
spectra; obviously, we could identify it because of high S/N ratio in
our data. The system is displaced by $\sim$400~\kms\ from system H. It
is thus likely that both those systems originate in objects from the
same group of galaxies.

\section{Notes on the variability of the absorbers}
\label{sec:variability}

Misawa et al.\ present convincing arguments that the system A absorber
is variable, but they did not see variability in any of the other
systems. In general, unlike the Subaru observations, our optical data
are not well suited for the analysis of absorber variability. First,
we have data from two epochs only. Second, the data were obtained with
two different telescopes/instruments, at different locations and under
different atmospheric conditions -- not to mention that one of the
observations was likely affected by smoke from forest fires. Third,
one of the extracted spectra had issues with residual flux. Fourth, in
some places the continuum fits had to be fudged manually.  Fifth, the
variability range reported in \citetalias{misawa2007} was of the order
of $\Delta\log N\sim0.3$, which is only slightly higher than our
typical measurement uncertainties. All those factors make direct
comparison of the derived properties of the absorption lines
questionable, and especially in the regions affected by the continuum
fit uncertainty. Any difference between absorber properties derived
from the two spectra will likely contain an unknown component
attributable to those issues.

\begin{figure}
\centering
\resizebox{\hsize}{!}{\includegraphics{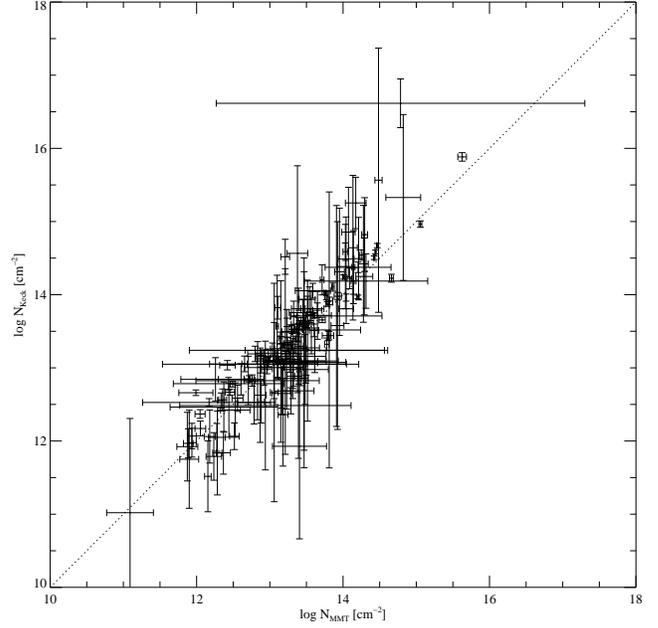}}
\caption{Comparison of column densities of all lines for which
measurements exist from both MMT (x-axis) and Keck (y-axis) data. See
text for discussion.}
\label{fig:mmtvskeck}
\end{figure}

Being aware of all those considerations, we examined the column
densities derived from the MMT and Keck data looking for any
systematic differences between the two datasets. In
Figure~\ref{fig:mmtvskeck} we show column densities for the same
components as derived from the MMT and Keck data. Note that the error
bars only show formal uncertainties as derived from the fits and do
not include the unknown uncertainty related to the analysis
issues. Despite that fact, one can clearly see that the properties of
components as derived from our two observations agree very well. Only
very few data points deviate from the $N_{\rm Keck}=N_{\rm MMT}$ line
by more than 1$\sigma$, and the ones that do are practically all in
the high-$N$ region, where issues related to the residual flux in the
MMT spectrum are expected to play a role.

This conclusion applies to both $\zabs\approx\zem$ -- including
system~A, which Misawa et al.\ have shown to be variable -- as well as
the $\zabs\ll\zem$ systems. For the latter systems this is not
surprising, although, as mentioned earlier, there were claims in the
literature \citep[e.g.][]{hamann1997,richards1999} indicating that
systems with large redshift difference showed variability \citep[see
also][]{hao2007}.

\section{Summary}
\label{sec:summary}

The following summarises our findings:

\begin{itemize}

\item The intrinsic X-ray properties of \HS\ are in the typical
range. The available X-ray data were insufficient to provide
meaningful constraints on the properties of the intrinsic
absorber. The optical-to-X-ray slope,  $\aox=1.70$, which is
at the high end of the typical range at this redshift. 

\item In addition to systems seen in \citetalias{dobrzycki1999} and
Misawa et al.'s papers, we identify two previously unknown intervening
systems at $\zabs=1.9266$, and 2.2622. With eleven complexes of heavy
element absorbers, which together contain $\sim$50 (or even more)
individual clouds -- and that not counting \Lya\ forest systems --
\HS\ is truly a spectacular object.

\item Some of the clouds in system A have a high ratio of \CIV\ to
\HI\ column densities, reaching $\sim$20.  Clouds from other
systems show $N_{\CIV}/N_{\HI}<1$. 

\item The data presented in this paper are not well suited for proper
analysis of line temporal variability. We note, however, that the data
are consistent with absence of variability for all systems.

\item The optical data presented in this paper contain identifications
and measurements of several lines not seen in the Subaru spectra, in
particular the hydrogen and magnesium lines.

\item Our data contain measurements of column densities of several
ions and therefore provide excellent constraints for modeling of the
cloud internal properties and QSO intrinsic flux. We will present such
modeling in the forthcoming paper.

\end{itemize}

\begin{acknowledgements}

We thank J.~Liske for his help with the VPGUESS/VPFIT packages,
T.~Aldcroft and A.~Siemiginowska for helpful comments, and the referee
for a very careful review. MN would like to thank ESO for hospitality
during his stay there. This work was partially supported by the ESO
Director General Discretionary Fund, by the National Aeronautics and
Space Administration through Chandra Award Number 04700527 issued by
the Chandra X-ray Observatory Center, which is operated by the
Smithsonian Astrophysical Observatory for and on behalf of the
National Aeronautics Space Administration under contract NAS8-03060,
and by grant 1P03D00829 of the Polish State Committee for Scientific
Research.

\end{acknowledgements}

\bibliographystyle{aa}
\bibliography{myrefs}

\onecolumn

\begin{longtable}{lccccc}
\caption{Lines from system A ($\zabs=2.44$) detected in the Keck spectrum.
\label{tab:sysa}} \\
\hline\hline
No.Line ID & \zabs & $b$             & $\log N$       & \Cf & M05 ID \\ 
           &       & (\kms)          & (\cmsq)        &             &           \\
\endfirsthead
\caption{continued.} \\
\hline\hline
No.Line ID & \zabs & $b$             & $\log N$       & \Cf     & M05 ID \\
           &       & (\kms)   & (\cmsq)        &         &        \\
\hline\hline
\endhead
\hline\hline

\hline
1.\HI & 2.41941\err0.00002 & 62.58\err6.51 & 13.656\err0.033 & 0.61 &	\\
1.\CIV\ \P & 2.41941\err0.00002 & 72.74\err3.42 & 14.965\err0.044 & 0.36 & 1; \CIV \\
1.\NV\ \sh& 2.41941\err0.00002 & 26.74\err17.05 & 13.235\err0.964 &&		\\
1.\SiIV & 2.41941\err0.00002 & 6.50 (fixed) & 12.065\err0.182 &&		\\

\hline
2.\HI\ \P & 2.42001\err0.00002 & 25.00 (fixed) & 14.496\err0.087 & 0.61 & \\
2.\CIV\ \P & 2.42001\err0.00002 & 198.64\err21.32 & 14.225\err0.054 & 0.36 & \\
2.\NV\ \sh& 2.42001\err0.00002 & 9.30 (fixed) & 13.008\err0.257 &&	\\
2.\SiIV & 2.42001\err0.00002 & 6.50 (fixed) & 11.789\err0.325 && 	\\

\hline
3.\HI  & 2.42344\err0.00025 & 25.00 (fixed) & 13.132\err0.037 & 0.61 & \\
3.\CIV\ \P & 2.42344\err0.00025 & 61.09\err18.07 & 13.440\err0.068 & 0.36 & \\
3.\NV\ \sh& 2.42344\err0.00025 & 9.30 (fixed) & 12.683\err0.319 &&	\\
\hline
4.\HI  & 2.42447\err0.00019 & 25.00 (fixed) & 13.131\err0.037 & 0.61 & \\
4.\CIV & 2.42447\err0.00019 & 34.72\err17.17 & 13.090\err0.333 & 0.36 & 2; \CIV \\
4.\NV\ \sh& 2.42447\err0.00019 & 9.30 (fixed) & 13.296\err0.349 && 	\\
\hline
5.\HI & 2.42554\err0.00002 & 25.00 (fixed) & 12.420 (fixed) & 0.61 & \\
5.\CIV\ & 2.42554\err0.00002 & 10.00 (fixed) & 14.313\err0.870 & 0.36 & 3; \CIV \\
5.\NV\ \sh& 2.42554\err0.00002 & 9.30 (fixed) & 12.885\err0.106 && 	\\
5.\SiIII\ $\sharp$& 2.42554\err0.00002 & 6.50 (fixed) & 12.667\err0.041 && \\
\hline
6.\HI\ \P & 2.42696\err0.00003 & 25.00 (fixed) & 14.700\err0.050 & 0.61 & 	\\
6.\CIV & 2.42696\err0.00003 & 44.13\err10.52& 13.322\err0.086 & 0.36 & 	\\
6.\NV\ \sh& (a) &&&&	\\
\hline
7.\HI  & 2.42833\err0.00001 & 25.00 (fixed) & 13.941\err0.016 &&		\\
7.\CIV & 2.42833\err0.00001 & 10.00 (fixed) & 13.526\err0.055 &&		\\
7.\NV\ \sh& 2.42833\err0.00001 & 9.30 (fixed) & 12.650\err0.200 && 	\\
\hline
8.\HI  & 2.43509\err0.00006 & 49.08\err30.79 & 12.361\err0.704 & & 	\\
8.\CIV & 2.43509\err0.00006 & 38.64\err8.37 & 13.440\err0.058 && 4; \CIV \\
8.\NV\ \sh& 2.43509\err0.00006 & 9.30 (fixed) & 12.977\err0.477 && 	\\
8.\SiIII\ \sh& 2.43509\err0.00006 & 6.50 (fixed) & $\le$16.000 && 	\\
\hline
9.\HI  & 2.43710\err0.00005 & 100.56\err33.64 & 13.091\err0.110 & & 	\\
9.\CIV & 2.43710\err0.00005 & 81.40\err8.82 & 13.970\err0.024 && 5; \CIV \\
9.\NV\ \sh& 2.43710\err0.00005 & 9.30 (fixed) & 13.094\err1.220 && 	\\
9.\SiIII\ \sh\ & 2.43710\err0.00005 & 6.50 (fixed) & $\le$16.000 && 		\\
9.\SiIV        & 2.43710\err0.00005 &  6.50 (fixed) & 12.050\err0.190 && \\

\hline
10.\HI  & 2.43753\err0.00002 & 25.00 (fixed) & 13.262\err0.061 && 	\\
10.\CIV & 2.43753\err0.00002 & 10.00 (fixed) & 13.596\err0.121 && 6; \CIV \\
10.\NV\ \sh& 2.43753\err0.00002 & 9.30 (fixed) & 13.717\err0.327 & & 	\\
10.\SiIII\ \sh& 2.43753\err0.00002 & 6.50 (fixed) & $\le$16.000 && 		\\
10.\SiIV & 2.43753\err0.00002 &  6.50 (fixed) & 11.839\err0.292 && \\

\hline
11.\HI  & 2.43911\err0.00030 & 138.45 (fixed) & 13.244\err0.087 && 	\\
11.\CIV & 2.43911\err0.00030 & 50.59\err40.10 & 13.273\err0.481 && 7; \CIV \\
11.\NV\ \sh& (a) &&&& \\
11.\SiIII\ \sh& 2.43911\err0.00030 & 6.50 (fixed) & $\le$ 15.376 && 	\\
\hline
12.\HI  & 2.43989\err0.00019 & 25.00 (fixed) & 12.464\err0.344 && 	\\
12.\CIV & 2.43989\err0.00019 & 10.00 (fixed) & 12.825\err0.534 && 8; \CIV \\
12.\NV\ \sh& 2.43989\err0.00019 & 9.30 (fixed) & 12.485\err0.880 && 	\\
12.\SiIII\ \sh\ & 2.43989\err0.00019 & 6.50 (fixed) & $\le$16.000 && 	\\
\hline
13.\HI  & 2.44079\err0.00042 & 72.42 (fixed) & 13.057\err0.103 && 	\\
13.\CIV & 2.44079\err0.00042 & 77.05\err50.41 & 13.518\err1.886 && 9; \CIV \\
13.\NV\ \sh& (a) &&&& \\
13.\SiIII\ \sh& 2.44079\err0.00042 & 6.50 (fixed) & $\le$12.734 && 	\\
\hline
14.\HI  & 2.44215\err0.00030 & 25.00 (fixed) & 13.004\err0.131 && 	\\
14.\CIV & 2.44215\err0.00030 & 10.00 (fixed) & 13.787\err0.100 && 10; \CIV \\
14.\NV\ \sh& 2.44215\err0.00030 & 9.30 (fixed) & 14.054\err0.101 && 	\\
14.\SiIII\ \sh& 2.44215\err0.00030 & 6.50 (fixed) & $\le$11.587 && 	\\
\hline
15.\HI  & 2.44228\err0.00021 & 81.41\err43.00 & 13.106\err0.221 &&	\\
15.\CIV & 2.44228\err0.00021 & 123.22\err87.76 & 13.240\err0.305 && 11; \CIV \\
15.\NV\ \sh& 2.44228\err0.00021 & 9.30 (fixed) & 13.808\err0.171 & & 	\\
15.\SiIII\ \sh& 2.44228\err0.00021 & 6.50 (fixed) & $\le$11.520 &&	\\
\hline
16.\HI  & 2.44325\err0.00148 & 114.39 (fixed) & 13.086\err0.203 && 	\\
16.\CIV & 2.44325\err0.00148 & 100.74 (fixed) & 13.709\err0.055 && 12; \CIV \\
16.\NV\ \sh& (a) &&&& 					\\
16.\SiIII\ \sh& 2.44325\err0.00148 & 6.50 (fixed) & $\le$12.394 & & 	\\
\hline
17.\HI  & 2.44480\err0.00018 & 91.30\err64.73 & 13.093\err0.104 && 	\\
17.\CIV & 2.44480\err0.00018 & 30.00\err20.54 & 13.070\err0.176 && 14; \CIV \\
17.\NV\ \sh& 2.44480\err0.00018 & 9.30 (fixed) & 13.094\err0.433 && 	\\
17.\SiIII\ \sh& 2.44480\err0.00018 & 6.50 (fixed) & $\le$12.679 && 	\\
\hline
18.\HI  & 2.44641\err0.00021 & 75.27\err25.93 & 13.154\err0.216 && 	\\
18.\CIV & 2.44641\err0.00021 & 109.27\err58.99 & 13.807\err0.202 && 15; \CIV \\
18.\NV\ \sh& (a) &&&& \\
18.\SiIII\ \sh& 2.44641\err0.00021 & 6.50 (fixed) & $\le$13.483 &&	\\
\hline
19.\HI  & 2.44818\err0.00002 & 25.00 (fixed) & 13.594\err0.020 && 	\\
19.\CIV & 2.44818\err0.00002 & 41.51\err49.68 & 12.920\err0.366 &&	\\
19.\NV\ \sh& (a) &&&& \\
19.\SiIII\ \sh& 2.44818\err0.00002 & 6.50 (fixed) & $\le$12.467 && 	\\
\hline
20.\HI  & 2.44954\err0.00035 & 41.27 (fixed) & 12.608 (fixed) && 	\\
20.\CIV & 2.44954\err0.00035 & 62.77\err55.72 & 13.133\err0.388 && 	\\
20.\NV\ \sh& (a) &&&& \\
20.\SiIII\ \sh& 2.44954\err0.00035 & 6.50 (fixed) & $\le$12.644 && 	\\
\hline
21.\HI  & 2.45069\err0.00082 & 25.00 (fixed) & 12.874 (fixed) &&	\\
21.\CIV & 2.45069\err0.00082 & 65.13\err108.68& 13.117\err0.681 &&	\\
21.\NV\ \sh& 2.45069\err0.00082 & 9.30 (fixed) & 11.928\err1.265 & & 	\\
21.\SiIII\ \sh& 2.45069\err0.00082 & 6.50 (fixed) & $\le$12.511 &&	\\
\hline
22.\HI\ \P& 2.45174\err0.00041 & 51.00 (fixed) & 14.245 (fixed) &&	\\
22.\CIV & 2.45174\err0.00041 & 53.88\err44.25 & 13.059\err0.340 && 	\\
22.\NV\ \sh& 2.45174\err0.00041 & 9.30 (fixed) & 12.686\err0.923 &&	\\
22.\SiIII\ \sh& 2.45174\err0.00041 & 6.50 (fixed) & $\le$14.411 && 	\\
\hline\hline
\end{longtable}
\noindent 
$\sharp$ -- heavy-element line in the \Lya\ forest region of the spectrum; \\
\P\ -- saturated line; \\
(a) -- detected in the MMT spectrum, but not seen in the Keck spectrum.

\clearpage

\begin{longtable}{lccccccc}
\caption{Lines from system B ($\zabs=2.48$) detected in the Keck spectrum.
\label{tab:sysb}} \\
\hline\hline
No.Line ID & \zabs & $b$& $\log N$ & M05 ID \\
   &  & (\kms) & (\cmsq) & \\

\endfirsthead
\caption{continued.} \\
\hline\hline
No.Line ID & \zabs & $b$ & $\log N$ & M05 ID \\
   & & (\kms) & (\cmsq) & \\
\hline\hline
\endhead
\hline\hline
1.\HI\ \P & 2.47723\err0.00002 & 25.00 (fixed) & 14.583\err0.228 &	\\
1.\CIV & 2.47723\err0.00002 & 10.00 (fixed) & 14.223\err0.141 & 1,2; \CIV\\
1.\CII & 2.47723\err0.00002 & 10.00 (fixed) & 13.320\err0.093 & 1,\CII \\
1.\NV\ \sh& 2.47723\err0.00002 & 9.30 (fixed) & 13.617\err0.426 &	\\
1.\MgII & 2.47723\err0.00002 & 7.00 (fixed) & 12.123\err0.300 &	\\
1.\AlIII & 2.47723\err0.00002 & 6.70 (fixed) & 11.923\err0.469 &	\\
1.\AlII & 2.47723\err0.00002 & 6.70 (fixed) & 11.020\err1.288 &	\\
1.\SiIV & 2.47723\err0.00002 & 6.50 (fixed) & 13.258\err0.082 & 1,2; \SiIV\\
1.\SiIII\ \sh& 2.47723\err0.00002 & 6.50 (fixed) & 13.068\err1.434 &	\\
1.\SiII & 2.47723\err0.00002 & 6.50 (fixed) & 12.554\err0.103 & 1,2; \SiII\\
\hline
2.\HI\ & 2.47762\err0.00004 & 25.00 (fixed) & 14.000\err0.154 &	\\
2.\CIV & 2.47762\err0.00004 & 10.00 (fixed) & 13.643\err0.110 & 3-5; \CIV\\
2.\NV\ \sh& 2.47762\err0.00004 & 9.30 (fixed) & 12.626\err0.645 &	\\
2.\MgII & 2.47762\err0.00004 & 7.00 (fixed) & 11.915\err0.416 & 	\\
2.\AlIII & (a) &&&	\\
2.\AlII & (a) &&&	\\
2.\SiIII\ \sh& 2.47762\err0.00004 & 6.50 (fixed) & 16.616\err0.333 & 1; \SiIII\\
2.\SiII & (a) &&&	\\
\hline						     
3.\HI\ \P & 2.47847\err0.00001 & 25.00 (fixed) & 14.625\err0.100 &	\\
3.\CIV & 2.47847\err0.00001 & 10.00 (fixed) & 15.265\err0.135 & 6-10; \CIV\\
3.\CII & (a) && 							 & 2; \CII \\
3.\NV\ \sh& 2.47847\err0.00001 & 9.30 (fixed) & 13.493\err0.072 & 1,2; \NV\\

3.\AlIII & (a) &&&	\\
3.\AlII & (a) &&&	\\
3.\SiIV & 2.47847\err0.00001 & 26.40\err3.23 & 12.827\err0.077 & 3,4; \SiIV\\
3.\SiIII\ \sh& 2.47847\err0.00001 & 26.40\err3.23 & 14.203\err0.206 & 2; \SiIII\\
3.\SiII & (a) &&	 					 & 3,4; \SiII\\
\hline						     
4.\HI\ \P & 2.47938\err0.00001 & 25.00 (fixed) & 15.810\err1.099 &	\\
4.\CIV & 2.47938\err0.00001 & 29.97\err1.36 & 14.515\err0.026 & 11-13; \CIV\\
4.\CII & 2.47938\err0.00001 & 29.97\err1.36 & 14.671\err0.029 & 3,4; \CII\\
4.\NV\ \sh,$\dag$& 2.47938\err0.00001 & 38.37\err30.95& 13.512\err0.087 & 3,4,5; \NV\\
4.\OI\ $\ddag$& 2.47938\err0.00001 & 8.70 (fixed) & 13.754\err0.184 & 1; \OI\\
4.\MgII\ $\dag$& 2.47938\err0.00001 & 7.00 (fixed) & 15.096\err0.159 &\\

4.\MgI & 2.47938\err0.00001 & 7.00 (fixed) & 11.766\err0.209 &		\\
4.\AlIII & 2.47938\err0.00001 & 6.70 (fixed) & 13.054\err0.107 &		\\
4.\AlII & 2.47938\err0.00001 & 6.70 (fixed) & 14.518\err0.239 & 1,2; \AlII\\
4.\SiIV & 2.47938\err0.00001 & 21.87\err5.96 & 13.398\err0.035 & 5,6; \SiIV\\
4.\SiIII\ \sh& 2.47938\err0.00001 & 21.87\err5.96 & 13.810\err0.120 & 3; \SiIII\\
4.\SiII & 2.47938\err0.00001 & 21.87\err5.96 & 13.938\err0.151 & 5,6; \SiII\\
4.\FeIII\ \sh$\dag$& (b) & & &						\\
4.\FeII\ \sh& 2.47938\err0.00001 & 4.65 (fixed) & 13.250 (fixed) & 1; \FeII	\\
\hline	
5.\HI\ \P & 2.47998\err0.00002 & 25.00 (fixed) & 14.110\err0.503 &		\\
5.\CIV & 2.47998\err0.00002 & 10.00 (fixed) & 14.550\err0.080 & 14,15; \CIV\\
5.\NV\ \sh& 2.47998\err0.00002 & 9.30 (fixed) & 12.843\err0.296 & 6;  \NV\\

5.\AlIII & 2.47998\err0.00002 & 6.70 (fixed) & 11.836\err0.573 &		\\
5.\AlII & 2.47998\err0.00002 & 6.70 (fixed) & 11.517\err0.484 & 3; \AlII\\
5.\SiIV & 2.47998\err0.00002 & 34.32\err5.42 & 13.075\err0.067 & 7; \SiIV\\
5.\SiIII\ \sh& 2.47998\err0.00002 & 34.32\err5.42 & 12.984\err0.065 & 4; \SiIII\\
5.\SiII & 2.47998\err0.00002 & 34.32\err5.42 & 11.355\err1.345 & 7; \SiII\\

5.\FeIII\ \sh& (b) & & &							\\
5.\FeII\ \sh& 2.47998\err0.00002 & 4.65 (fixed) & 12.843\err0.214 &		\\
\hline    
6.\HI\ \P & 2.48064\err0.00002 & 25.00 (fixed) & 14.969\err0.590 &		\\
6.\CIV & 2.48064\err0.00002 & 10.00 (fixed) & 14.110\err0.076 & 16,17; \CIV\\
6.\CII & 2.48064\err0.00002 & 10.00 (fixed) & 13.983\err0.053 & 6,7; \CII\\
6.\NV\ \sh& 2.48064\err0.00002 & 9.30 (fixed) & 12.423\err0.371 &		\\
6.\MgII & 2.48064\err0.00002 & 7.00 (fixed) & 12.903\err0.115 &		\\
6.\AlIII & 2.48064\err0.00002 & 6.70 (fixed) & 12.061\err0.362 &		\\
6.\AlII & 2.48064\err0.00002 & 6.70 (fixed) & 11.968\err0.199 & 4; \AlII\\
6.\SiIV & 2.48064\err0.00002 & 6.50 (fixed) & 13.334\err0.094 & 8,9; \SiIV\\
6.\SiIII\ \sh,$\dag$& 2.48064\err0.00002 & 6.50 (fixed) & 15.883\err0.058 & 5-7; \SiIII\\
6.\SiII & 2.48064\err0.00002 & 6.50 (fixed) & 13.136\err0.122 & 8; \SiII\\
6.\FeIII\ \sh& (b) & & &							\\
\hline
7.\HI\ \P & 2.48145\err0.00002 & 25.00 (fixed) & 14.431\err0.338 &		\\
7.\CIV & 2.48145\err0.00002 & 10.00 (fixed) & 13.979\err0.072 & 18,19;\CIV\\
7.\CII & 2.48145\err0.00002 & 10.00 (fixed) & 13.914\err0.052 & 8; \CII\\
7.\NV\ \sh& 2.48145\err0.00002 & 9.30 (fixed) & 12.423\err0.371 &	\\
7.\MgII & 2.48145\err0.00002 & 7.00 (fixed) & 13.027\err0.112 & 	\\
7.\AlIII & 2.48145\err0.00002 & 6.70 (fixed) & 12.562\err0.150 &	\\
7.\AlII & 2.48145\err0.00002 & 6.70 (fixed) & 12.068\err0.174 & 5; \AlII\\
7.\SiIV & 2.48145\err0.00002 & 6.50 (fixed) & 13.407\err0.082 & 10; \SiIV\\
7.\SiIII\ \sh& 2.48145\err0.00002 & 6.50 (fixed) & 13.936\err0.171 & 8; \SiIII\\
7.\SiII & 2.48145\err0.00002 & 6.50 (fixed) & 13.449 (fixed) & 9; \SiII\\
7.\FeIII\ \sh& (b) & & &							\\
7.\FeII\ \sh& 2.48145\err0.00002 & 4.65 (fixed) & 12.526\err0.293 &		\\
\hline						     
8.\HI\ \P & 2.48226\err0.00002 & 25.00 (fixed) & 15.111\err0.100 &		\\
8.\CIV & 2.48226\err0.00002 & 10.00 (fixed) & 12.537\err0.407 &		\\
8.\CII\ $\dag$& 2.48226\err0.00002 & 10.00 (fixed) & 14.374\err0.072 & 9-11; \CII\\
8.\NV\ \sh& (a) &&& \\
8.\OI  & 2.48226\err0.00001 & 8.70 (fixed) & 13.549\err0.066 &   \\
8.\MgII & 2.48226\err0.00002 & 7.00 (fixed) & 13.014\err0.106 & 		\\
8.\AlIII & 2.48226\err0.00002 & 6.70 (fixed) & 11.751\err0.670 &		\\
8.\AlII & 2.48226\err0.00002 & 6.70 (fixed) & 11.977\err0.197 & 6,7; \AlII\\
8.\SiIV & (a) &&&	 					\\
8.\SiIII\ \sh& 2.48226\err0.00002 & 6.50 (fixed) & 13.124\err0.113 & 9,10; \SiIII\\
8.\SiII & 2.48226\err0.00002 & 6.50 (fixed) & 13.822\err0.155 & 10,11; \SiII\\

\hline\hline
\end{longtable}
\noindent
$\sharp$ -- heavy-element line in the \Lya\ forest region of the spectrum; \\
$\ddag$ -- line blended by identified line (\OI\ line is blended by
\CIV\ from system J);\\ 
$\dag$ -- line is probably blended with an unidentified line;\\
\P\ -- saturated line; \\
(a) -- detected in the MMT spectrum, but not seen in the Keck spectrum;\\
(b) -- detected in the MMT spectrum, but noise is too strong in
order to fit this line in the Keck spectrum.

\clearpage

\begin{longtable}{lcccc}
\caption{Lines from systems C and D detected in the Keck spectrum.
\label{tab:syscd}} \\
\hline\hline
No.Line ID & \zabs & $b$ & $\log N$ & M05 ID \\
   & & (\kms) & (\cmsq) & \\
\endhead
\hline\hline
\multicolumn{5}{c}{System C $\zabs = 2.54$} \\
\hline
1.\HI\ \P& 2.53787\err0.00003 & 25.00 (fixed) & 15.659\err0.270	&	\\
1.\CIV & 2.53787\err0.00003 & 8.32\err4.25 & 13.510\err0.155	& 1; \CIV\\
\hline\hline

\multicolumn{5}{c}{System D $\zabs = 2.55$} \\
\hline
1.\HI & 2.55222\err0.00001 & 25.00 (fixed) & 13.569\err0.697 &	\\
1.\CIV & 2.55222\err0.00001 & 10.00 (fixed) & 12.585\err0.410 & 1; \CIV\\
\hline						     
2.\HI\ & 2.55383\err0.00015 & 25.00 (fixed) & 14.414\err1.474 &	\\
2.\CIV & 2.55383\err0.00015 & 10.00 (fixed) & 13.456\err0.790 &	2,\CIV\\
\hline						     
3.\HI\ \P& 2.55409\err0.00018 & 25.00 (fixed) & 16.042\err1.916 & \\
3.\CIV \P& 2.55409\err0.00018 & 14.45\err7.54 & 14.407\err0.863 & 3; \CIV\\
3.\SiIV& 2.55409\err0.00018 & 11.81\err8.00 & 12.850\err0.043 & 1; \SiIV\\
\hline						     
4.\HI\ \P& 2.55442\err0.00005 & 25.00 (fixed) & 14.865\err1.732 &	\\
4.\CIV \P& 2.55442\err0.00005 & 11.40\err3.96 & 15.200\err1.134 & 4; \CIV\\
4.\SiIV& 2.55442\err0.00005 & 10.64\err4.40 & 13.593\err0.085 & 2; \SiIV\\
4.\NV\ \# & 2.55442\err0.00005 & 38.57\err16.44& 13.191\err0.105 &	\\
\hline\hline
\end{longtable}
\noindent \P\ -- saturated line; \\
\#\ -- likely blend of components 3 and 4.

\clearpage

\begin{longtable}{lcccc}
\caption{Lines from systems E-K detected in the Keck spectrum.
\label{tab:sysek}}\\
\hline\hline
No.Line ID & \zabs & $b$ & $\log N$ & M05 ID \\
   & & (\kms) & (\cmsq) &\\
\hline\hline
\endfirsthead
\caption{continued.} \\
\hline\hline
No.Line ID & \zabs & $b$ & $\log N$ & M05 ID \\
   & & (\kms) & (\cmsq) &\\
\hline\hline
\endhead
\multicolumn{5}{c}{System E $\zabs = 1.89$} \\
\hline
1.\HI  & (a) &&&\\
1.\CIV  & 1.88805\err0.00001 & 10.00 (fixed) & 13.873\err0.032 & 1; \CIV\\
1.\SiIV\ \sh& 1.88805\err0.00001 & 6.50 (fixed) & 13.036\err0.064 &	\\

\hline
2.\HI  & (a) &&&\\
2.\CIV  & 1.88851\err0.00007 & 10.00 (fixed) & 13.576\err0.034 & 2; \CIV\\
2.\SiIV\ \sh & (a) & & & \\
\hline\hline
\multicolumn{5}{c}{System J (new system) $\zabs = 1.93$} \\
\hline
1.\HI   & (a) &&&\\
2.\CIV\ $\ddag$& 1.92658\err0.00001 & 10.00 (fixed) & 13.727\err0.081 &	\\

\hline\hline
\multicolumn{5}{c}{System F $\zabs = 1.97$} \\
\hline
1.\HI  & (a) &&&\\
1.\CIV & 1.96399\err0.00002 & 10.00 (fixed) & 13.699\err0.213 & 1; \CIV\\

\hline
2.\HI  & (a) &&&\\
2.\CIV & 1.96420\err0.00002 & 10.00 (fixed) & 14.185\err0.179 & 2,3; \CIV\\

\hline
3.\HI  & (a) &&&\\
3.\CIV & 1.96519\err0.00001 & 10.00 (fixed) & 14.241\err0.030 & 4,5; \CIV\\
\hline
4.\HI  & (a) &&&\\
4.\CIV & 1.96572\err0.00001 & 10.00 (fixed) & 13.574\err0.026 & 6; \CIV\\
\hline\hline
\multicolumn{5}{c}{System G $\zabs = 2.07$} \\
\hline
1.\HI  & (a) &&&\\
1.\CIV  & 2.07061\err0.00005 & 10.00 (fixed) & 14.385\err0.576 & 1; \CIV\\
1.\SiIV\ \sh& 2.07061\err0.00005 & 6.50 (fixed) & 12.887\err0.307 & 1; \SiIV\\

\hline
2.\HI  & (a) &&&\\
2.\CIV  & 2.07091\err0.00005 & 10.00 (fixed) & 14.487\err0.575 & 2; \CIV\\
2.\SiIV\ \sh& 2.07091\err0.00005 & 6.50 (fixed) & 13.491\err0.188 & 2; \SiIV\\
\hline\hline
\multicolumn{5}{c}{System I $\zabs = 2.18$} \\
\hline
1.\HI     & (c) & & &    \\
1.\CIV    & 2.17560\err0.00002 & 10.00 (fixed) & 13.295\err0.036 & \\
1.\SiIV   & 2.17560\err0.00002 &  6.50 (fixed) & 12.359\err0.069 & \\
\hline
2.\HI     & (c) & & &    \\
2.\CIV    & 2.17606\err0.00005 & 10.00 (fixed) & 13.066\err0.056 & \\
2.\SiIV   & 2.17606\err0.00005 &  6.50 (fixed) & 12.107\err0.123 & \\
\hline
3.\HI     & (c) & & &    \\
3.\CIV    & 2.17645\err0.00001 & 10.00 (fixed) & 13.741\err0.033 & \\
3.\SiIV   & 2.17645\err0.00001 &  6.50 (fixed) & 12.750\err0.040 & \\

\hline\hline
\multicolumn{5}{c}{System K (new system) $\zabs = 2.26$} \\
\hline
1.\HI\ \P & 2.26214\err0.00005 & 25.00 (fixed) & 14.850\err0.615 &	\\
1.\CIV & 2.26214\err0.00005 & 10.00 (fixed) & 12.827\err0.332 & 	\\

\hline\hline
\multicolumn{5}{c}{System H $\zabs = 2.27$} \\
\hline
1.\HI\ \P & 2.26613\err0.00002 & 25.00 (fixed) & 14.075\err0.570 &	\\

1.\CIV & 2.26613\err0.00002 & 10.00 (fixed) & 13.200\err0.035 & 1; \CIV\\
1.\SiIV & 2.26613\err0.00002 & 6.50 (fixed) & 12.527\err0.053 &	\\
1.\SiIII\ \sh& 2.26613\err0.00002 & 6.50 (fixed) & 12.784\err0.085 & 1; \SiIII\\

\hline\hline
\end{longtable}
\noindent 
$\sharp$ -- heavy-element line in \Lya\ forest region of the spectrum; \\ 
$\ddag$ -- line blended with identified line (\CIV\ in system J
is blended with \OI\ from system B); \\
\P\ -- saturated line; \\
(a) -- \HI\ lines are not observable in systems E, J, F, I and G, in
either MMT or Keck spectra. \HI\ lines are seen in system~I in the MMT spectrum.

\clearpage 

\begin{longtable}{lccc}
\caption{Total column densities of ions detected in the Keck spectrum.
\label{tab:logNtot}} \\
\hline\hline
Ion & $\log N_{\rm tot}$(this paper)& $\log N_{\rm tot}$(M05)& $\log N_{\rm tot}$(M07)\\
 & (\cmsq)     & (\cmsq)   & (\cmsq) \\
\hline\hline
\endfirsthead
\caption{continued.} \\
\hline\hline
Ion & $\log N_{\rm tot}$(this paper)&$\log N_{\rm tot}$(M05)& $log N_{\rm tot}$(M07)\\
 & (\cmsq)     & (\cmsq)   & (\cmsq) \\
\hline\hline
\endhead
\multicolumn{4}{c}{System A $\zabs = 2.44$} \\
\hline
\HI & $15.124 \pm 0.030$ & & \\
\CIV & $15.270 \pm 0.105$ & 15.236 & \\
\NV & $14.527 \pm 0.102$ & & \\
\SiIV & $12.250 \pm 0.164$ & & \\
\SiIII& (a) & & \\
\hline
\multicolumn{4}{c}{System B $\zabs = 2.48$} \\
\hline
\HI & $15.950 \pm 0.713$ & & \\
\CIV & $15.471\pm0.085$ & $15.024\pm0.047$ & $15.144\pm0.162$ \\
\CII & $14.956\pm0.025$ & $14.905\pm0.151$ & $15.570\pm0.997$ \\
\NV & $14.085\pm0.151$ & $13.663\pm0.275$ & $13.730\pm0.023$ \\
\OI & $13.965\pm0.116$ & $14.390\pm0.032$ & $14.608\pm0.028$ \\
\MgII & $15.106\pm0.155$ & & \\
~~~~~~~~($i$)& $13.493\pm0.062$ & & \\
\MgI & $11.766 \pm 0.209$ & & \\
\AlIII& $13.260\pm0.085$ & & \\
\AlII & $14.522\pm0.237$ & $12.964\pm0.023$ & $13.165\pm0.889$ \\
~~~~~~~~($ii$)& $12.542\pm0.113$ & & \\
\SiIV (2a)& $14.037\pm0.032$ & $13.909\pm0.020$ & $14.017\pm0.014$ \\
\SiIII& $16.693\pm0.279$ & $14.880\pm0.470$ & $15.328\pm0.301$ \\
~~~~~~~~($iii$)&$15.902\pm0.056$ & & \\
\SiII (3)&$14.298\pm0.084$ & $15.977\pm0.471$ & $15.783\pm0.354$ \\
\FeIII& (b) & & \\  
\FeII (4)& $13.449\pm0.064$ & $13.600\pm0.030$& $13.650\pm0.010$ \\
\hline
\multicolumn{4}{c}{System C $\zabs = 2.54$} \\
\hline
\HI & $15.659 \pm 0.270$ & & \\
\CIV & $13.510 \pm 0.155$ & $13.440 \pm 0.110$ & 
$13.410 \pm 0.050$ \\
\hline
\multicolumn{4}{c}{System D $\zabs = 2.55$} \\
\hline
\HI & $16.081\pm1.756$ & & \\
\CIV & $15.272\pm0.967$ & $14.919\pm0.200$ & $16.662\pm0.060$ \\
\NV & $13.191 \pm 0.105$ & & \\
\SiIV (2a)& $13.665\pm0.072$ & $13.514\pm0.091$ & $13.567\pm0.023$ \\
\hline\hline
\multicolumn{4}{c}{System E $\zabs = 1.89$} \\
\hline
\CIV & $14.050\pm0.024$ & $14.054\pm0.163$ & $14.034\pm0.074$ \\
\SiIV\ \sh & $13.036\pm0.064$ & & \\
\hline
\multicolumn{4}{c}{System J (new system) $\zabs = 1.93$} \\
\hline
\CIV\ $\ddag$ & $13.727\pm0.081$ & & \\
\hline
\multicolumn{4}{c}{System F $\zabs = 1.97$} \\
\hline
\CIV & $14.618\pm0.072$ & $15.840\pm3.410$ & $14.505\pm0.024$ \\
\hline
\multicolumn{4}{c}{System G $\zabs = 2.07$} \\
\hline
\CIV & $14.740\pm0.410$ & $14.227\pm0.054$ & $14.331\pm0.021$ \\
\SiIV (2b)& $13.587\pm0.162$ & $13.260\pm0.054$ & $13.327\pm0.017$ \\
\hline
\multicolumn{4}{c}{System I $\zabs = 2.18$} \\
\hline
\CIV & $13.928\pm0.023$ & & $14.019\pm0.072$ \\
\SiIV & $12.991\pm0.032$ & & $12.797\pm0.015$ \\
\SiIII& & & $12.800\pm0.022$ \\
\hline
\multicolumn{4}{c}{System K (new system) $\zabs = 2.26$} \\
\hline
\HI & $14.850\pm0.615$	& & \\
\CIV & $12.827\pm0.332$ & & \\
\hline
\multicolumn{4}{c}{System H $\zabs = 2.27$} \\
\hline
\HI &	$14.074\pm0.570$ & & \\
\CIV & $13.200\pm0.035$ & $13.230\pm0.330$ & $13.914\pm0.141$ \\
\SiIV & $12.527\pm0.053$ & & \\
\SiIII& $12.784\pm0.085$ & $13.020\pm2.410$ & $12.807\pm0.119$ \\
\hline\hline
\end{longtable}
\noindent M05 data taken from their Tables 1 and 2 (second
spectrum in both cases). M07 data taken from their Table 3.\\
The symbols denote:\\

($i$) -- without blended line 4.\MgII\ in system B;\\
($ii$) -- without blended line 4.\AlII\ in system B;\\
($iii$) -- without blended line 2.\SiIII\ in system B;\\
(a) -- but ion detected, but $N$ measurements are unreliable due to blends;\\
(b) -- \FeIII\ in system~B are detected in the MMT spectrum, but
noise is too strong in order to fit this line in the Keck spectrum.\\
$\sharp$ -- heavy-element line in \Lya\ forest region of the spectrum;\\
$\ddag$ -- blend.

\end{document}